\documentstyle[epsfig]{l-aa}
\begin{document}

\def\be{\begin{equation}}
\def\ee{\end{equation}}
\def\bm#1{\hbox{\boldmath $#1$}}

\thesaurus{ 02.02.1; 02.08.1; 03.13.4}
 \title{Numerical study of the tidal interaction of a star and a massive
black hole}
\author{J.A. Marck${}^\dagger$, A. Lioure${}^\ddagger$
			and S. Bonazzola${}^\dagger$ }
\offprints {J.A. Marck, marck@obspm.fr}
\institute {
 ${}^\dagger$  D\'epartement d'Astrophysique Relativiste et de Cosmologie,
Observatoire de Paris, section de Meudon,
 (UPR 176 du C.N.R.S.),
 92195 Meudon Cedex, France \\
 ${}^\ddagger$ Centre d'Etudes de Bruy\`eres-le-Ch\^atel,
Service P.T.N.,
 B.P. 12,
91680 Bruy\`eres-le-Ch\^atel, France.}
\date{Received 1994; accepted }

\maketitle
\markboth{Marck et al.:}{ star-black hole tidal interaction}
\begin{abstract}
We present a formalism well adapted to the numerical study of
the encounter of an ordinary
main sequence star with a massive black hole. Symmetry considerations,
the use of a
well adapted moving  grid and a well adapted moving frame
along with integration of the partial differential equations
by means of pseudo-spectral methods result in a very powerful and
accurate tool.
The hydrodynamical equations are written in a
moving frame which mimics the bulk of the movement of the fluid,
resulting in very small relative velocities and a well suited spatial
resolution throughout the calculations.
Therefore, the numerical
calculations are considerably simplified,
smoothing in particular the Courant condition. Typical runs are performed
within
a few hours on a workstation with the high accuracy linked to the spectral
methods.
Predictions of the so-called affine are tested against this full numerical
simulation.


\end{abstract}

\section {Introduction}
\bigskip

The problem of the tidal influence of a massive black hole on an
ordinary star is of great astrophysical interest, particularly in the
case of close encounters. Carter and Luminet (1982, 1983) already
suggested that the deep penetration of a star within the Roche radius
of a black hole should strongly perturb its core  and result in
metal-enriched winds flowing out of the black hole, or even in helium
detonation.  Their model is a semi-analytical treatment based on the
assumption of a linearized Lagrangian motion of the fluid. As a
consequence, the shape of the star remains ellipsoidal.  This model is
referred to as the affine model.  It was numerically exploited in
Luminet \& Carter (1986) to predict the central density, temperature
and entropy increases and in Luminet \& Pichon (1989a,b) to
estimate the additional nuclear energy release and the corresponding
production of heavy elements. They showed in particular that, contrary
to previous expectations, helium detonation by the triple-alpha
proccess could probably not occur although proton and alpha captures
could change significantly the chemical composition of the star.

A new motive for interest in this subject has been provided by recent
theoretical considerations by Carter (1992) which show that the tidal
disruption of an ordinary main sequence star is a conceivable scenario
for the gamma ray bursts.  Carter argues that the available energy that
might be radiated away through gamma rays, if a suitable transfert
mechanism were available, would be of the order of the initial binding
energy of the star, provided that the encounter is sufficently deep,
with penetration factors of the order of 10. Such a mechanism might
arise from an unstable shock formation due to deviations from affine
behaviour.

A qualitative description of the encounter is possible by making
further approximations, depending on which part of the track the star
is orbiting. One can split the movement of the star around the black
hole into five distinct phases. The first two phases correspond to
fairly clear and reliable approximations. First, far from the black
hole, the star is supposed to be in rough hydrostatic equilibrium. When
entering the Roche radius, the tidal acceleration tends to be dominant
compared to self-gravity. The next three phases are the bounce, where
pressure terms take over, then a rebound, which is an expansion with
only tidal acceleration, and finally an ejection of material, possibly
driven by nuclear energy release. During these last three phases, the
affine model becomes more and more unrealistic for several reasons:

first, the geometry deviates strongly from an exact ellipsoid when the
star takes a double-wedged shape due to the wringer effect of the tidal
field (see e.g. Bicknell and Gingold 1983 or Carter 1992);

second, hydrodynamical effects, like shocks, are likely to occur during
the phase of very strong and rapid compression;

third, if nuclear reactions are to be enhanced by the high temperatures
achieved, or if a significant fraction of the total energy is released
by electromagnetic waves (Carter 1992), the polytropic approximation
becomes unrealistic.

\noindent However, the affine model performs a very accurate
description of the movement of the star during the first two phases.
\par

\bigskip

To describe in more details the high compression phase, it then appears
interesting to develop a hydrodynamical code which would be able to
accurately follow the evolution of the star near the periastron and
after, in order to give quantitative results on the possible generation
of shocks and detonation of nuclear reactions. Several numerical
investigations have tried to go beyong the affine treatment in working
out the real deformation of the star near the black hole. The first
investigation of this kind was made by Bicknell and Gingold (1983),
using a 3--D Smoothed Particle Hydrodynamics (SPH) method.
 Their treatment was based on purely newtonian calculations.  Their
 main result concerned the maximum heating and maximum compression of
 the star: they found less dramatic effects than expected on the basis
 of the affine model and concluded that the triple-$\alpha $ could
 probably not detonate although CNO reactions could change
 significantly the chemical composition.  Morerecent SPH calculations
were made by Evans \& Kochanek, with a much better resolution but in
the axisymmetric approximation and not in full 3--D. Although their
treatment is fully relativistic, they are interested only in the debris
and not in the core itself, and therefore perform their calculation
with a relatively small penetration factor. Further very recent
 publications address  this problem again.  Khokhlov et al.
(1993a,b) report 3--D eulerian calculations where they examine the
energy and angular momentum tranfer to the star in order to check
whether it might be disrupted or not.  They do find a central density
increase although quantitatively different from the one  predicted by
Carter and Luminet (1983).
However, they mention that their numerical method might not be
 reliable in every case due to its poor resolution.  The same group
 (Frolov et al. 1994) followed up even more recently these calculations
 by including fully relativistic effects, both in the orbit and in the
 tidal field, using the analytic treatment by Marck (1983). They find
 quantitative differences from the non-relativistic case although the
 qualitative overall behaviour is similar. But the main effects
 described in those papers concern the outermost parts of the star,
 where the density is low, and the stripped mass does not exceed $10\%$
 of the total star mass.  Laguna et al. (1993) have carried out
 3--D relativistic SPH calculations to compute the evaporation of the
 star and the possible influence on emitted energy. They find
 significant deviations from the affine model which are compatible with
the results of Bicknell and Gingold (1983) for the maximum compression.
However, the SPH methods are known to be questionable (see e.g.
Hernquist 1993) since they effectively involve a high artificial
viscosity that might lead to considerable entropy production in high
compression phase  and thus give erroneous results for the maximum
central density which is crucial if one wants to determine what kind of
nuclear reactions can be initiated or not. In a fully realistic
description, one would expect energy dissipation in shocks, but it is
still unclear how much of it will occur. It may be conjectured, at this
stage, that the
 true outcome is intermediate between the predictions of the (strictly
 non dissipative) affine treatment and the (too highly dissipative) SPH
treatment.

\bigskip

The purpose of the present paper is to describe a new numerical
approach which should help to settle these questions. The essential idea
is to combine the use of a moving grid derived from the affine approach
with the very powerful and reliable pseudo-spectral methods that have
recently been developped for other purposes (Bonazzola \& Marck 1990).
The method we used is examplified with one typical run. This is
to be followed in the near future by an extensive study of the physical
aspects of the encounter.

A specific encouter may be characterized by a penetration factor
$\beta$. The definition of such a quantity is not unique. We choose the
following form for $\beta:$
\be
	\beta = R_R / R_P
\ee
where
\be
	R_R = R_\star {\left(M_h/M_\star \right)}^{1/3}
\ee
is the Roche radius, $R_P$ the periastron (minimum distance of the star
to the black hole), $R_\star$   a characteristic radius of the star and
$M_h$ and $M_\star$   the mass of the hole and the mass of the star
respectively. This definition agrees with the one taken by Laguna et
al. (1993) but differs from that of Khokhlov et al. (1993a): their
parameter $\eta$ can be identified with $\beta^{-3/2}.$ The details of
the method are given in Sections 2 and 3. Preliminary results are
presented in Section 4. Section 5 is devoted to discussion and
prospects. \par

\section {Description of the method}

The study of tidal interactions needs a full 3--D calculation. There
are two mains categories of approaches: the Eulerian and the Lagrangian
ones. In the first method, one writes the equations  of motion with
respect to a static frame. In the second method, one writes the
equations of motion in a frame comoving with the matter. The Eulerian
method can be easily worked out but would require prohibitive number of
mesh points to accurately describe an inhomogeneous distribution of
matter. The Lagrangian approach overcomes this drawback. However, in
the case of multidimentionnal hydrodynamics, one has generally to take
care of the formation of caustics. The Smoothed Particles Hydrodynamics
methods, which is a Lagrangian description, can be applied whatever the
matter distribution. However SPH is intrinsically highly dissipative
and, hence, may lead to inaccurate results especially in the study of
shock formation. \par

\bigskip

We introduce an intermediate approach which combines the advantages of
the Eulerian and Lagrangian methods. We solve the equations of motion
in a moving frame attached to the mean motion of the fluid. In the
particular case of tidal interaction of a star and a massive black
hole, the mean motion of the fluid is accurately described by the
affine star model of Carter \& Luminet (1983). After recalling the main
features of the affine model, we write down the exact hydrodynamical
equations in a general ellipsoidal coordinate system associated to its
canonical frame.

\subsection{The affine model}

The fundamental hypothesis of the affine model   described by Carter
and Luminet (1983) is that the position of each cell of fluid with
respect to a parallely propagated frame tied to the center of mass of
the star can be described by a linear lagrangian transformation:
\be
   r_{i}(t) = {\cal D}_{ij}(t) \hat{r}_j
\ee
where $\hat{\bm{r}}$ is the initial position vector of a fluid element
and $ \bm{ r}$ is the current position vector in a frame which is
parallely propagated along the orbit of the centre of mass of the star.
Hence, the unknowns are the nine coefficients ${\cal D}_{ij}$ of the
deformation matrix $\cal D,$ which satisfy a system of second order
differential equations which can be derived from a lagrangian. It turns
out that, within this formalism, the star keeps ellipsoidal
configurations. The principal axes of this ellipsoid are the
eigenvectors of the matrix ${\cal D}{}^t{\cal D}$. In the case of a
planar orbit (i.e. newtonian approximation or non-rotating black hole
or an orbit lying in the equatorial plane of the Kerr black hole) the
movement of the fluid in the $z$--direction and in the plane of the
orbit decouple. As a consequence, ${\cal D}$ has only 5 non-zero
components. Solving the equations of motion for ${\cal D}$ gives all
the information about the physics of the encouter within the
approximations made. The mass-density of each cell is given by
\be
	\rho ( \bm{r},t) = {\rho (\hat{\bm{r}} , 0) \over \vert {\cal D} \vert}
\ee
and the velocity is
\be
	\bm{v} = \dot{\cal D} \hat{\bm{r}} \ .
\ee
In the particular case of a polytropic equation of state $P \sim
\rho^\gamma$, the heat function of each cell obeys
\be
	h ( \bm{r},t) = {h (\hat{\bm{r}} , 0) \over \vert {\cal D} \vert^{\gamma-1}} \
{}.
\ee
When the star penetrates deeply inside the Roche radius, the surface of
its equatorial section remains approximatively constant while the star
undergoes a strong compression in the $z$--direction. This induces an
overall compression. The maximum value reached by the central density
is roughly given by:
\be
	\rho_m \sim \rho_0 \beta^{2 / (\gamma-1)} \ .
\ee
The typical duration of this high compression phase is
\be
	\tau_m \sim \beta^{ - {(\gamma+1)/(\gamma-1)} } \ .
\ee

\subsection{Hydrodynamics keeping the properties of the affine star model in
mind}

We describe in this subsection the method we used to build our
hydrodynamical code. Our point is to take advantage of most of the
analytical results coming out of the affine model by writing the
hydrodynamical equations in a well suited frame. We will be concerned
in this paper with polytropic fluids, obeying $P \sim \rho^\gamma.$
In that particular case, the enthalpy $h$ is the right energy variable
as shown in Marck and Bonazzola (1992) . The equations to be solved
write:

- mass conservation:
\be
	\partial_t \rho = -  \bm{\nabla \cdot} \left( \rho \bm{  v} \right)
\ee

- energy conservation:
\be
	\partial_t  h  =  - \bm{  v \cdot  \nabla} h
		- (\gamma -1) h  \bm{\nabla \cdot v}
\ee

- momentum conservation:
\be
\partial_t \bm{ v} = -\bm{ v \cdot  \nabla  v  -
			\nabla} (h +  \Phi  + {\cal C} )
\ee

- Poisson equation
\be
	\Delta \Phi = 4\pi G \rho
\ee
where ${\Phi}$ stands for the self-gravity potential and $\cal C$ for
the tidal potential. Note that the continuity equation is not necessary
when one uses an  adiabatic relation $P \sim \rho^\gamma.$

Let us now introduce the following coordinate transformation:
\be
\left(
	\begin{array}{c}
		\tau \\
		X^\prime
	\end{array}
\right)
			=
\left(
	\begin{array}{c c}
		1	&0 \\
		0	& {\cal Q}^{-1} (t)
	\end{array}
\right)
\left(
	\begin{array}{c}
		t \\
		X
	\end{array}
\right)
\ee
where $ X = \pmatrix{ x \cr y \cr z } $ are cartesian coordinates and
where ${\cal Q}(t)$ is some $3\times 3$ real regular matrix. Consider
now the Jacobian matrix ${\cal J}(t)$ associated with the previous
transformation:
\be
\left(
	\begin{array}{c}
		\partial _\tau   \\
		\partial _{x^\prime} \\
		\partial _{y^\prime} \\
		\partial _{z^\prime} \\
	\end{array}
\right)
  =  {~^t\cal J}
\left(
	\begin{array}{c}
		\partial _t   \\
		\partial _x \\
		\partial _y \\
		\partial _z \\
	\end{array}
\right)
\ee
where
\be
 {\cal J} =
\left(
	\begin{array}{c c}
		1	&0 \\
		\dot{\cal Q} {\cal Q}^{-1} X	& {\cal Q}
	\end{array}
\right)
\ee
and where a dot stands for time derivative.  The time derivative  operator
is corrected for the grid movement, and thus is suitable for
calculating relative velocity and acceleration with respect to the new
frame. The relative velocity of the fluid with respect to the moving
frame canonically associated to $(\tau,X^\prime)$ reads
\be
   {\bf \tilde v} = {\cal Q}^{-1} \bm{ v } -
		{\cal Q}^{-1} \dot{{\cal Q}}  \bm{ X^\prime }
\ee
and the hydrodynamical equations written in this new frame become

- mass conservation:
\be \label{eqr}
  \partial_{\tau} \rho = - {\bf \tilde \nabla \cdot} \left( \rho {\bf \tilde v}
\right) -
 			 \rho \partial_\tau \log \vert {\cal Q} \vert
\ee

- energy conservation:
\be \label{eqh}
\partial_{\tau}  h  =  -{ \bf \tilde v \cdot \tilde \nabla} h -
			(\gamma -1) h  \left(
				{\bf \tilde \nabla \cdot  \tilde v}  +
 		 		\partial_\tau \log \vert {\cal Q} \vert
					\right)
\ee

- momentum conservation:
\begin{eqnarray}
\partial_{\tau} {\bf \tilde v = - \tilde v \cdot \tilde \nabla \tilde v}
			&-& 2{\cal Q}^{-1} \dot{\cal Q} {\bf \tilde v}
			- {\cal Q}^{-1} \ddot{\cal Q} \bm{X^\prime} \\ \nonumber
			&-& {\cal Q}^{-1}~^t{\cal Q}^{-1}
			{\bf {\tilde\nabla} } \left( h +   \Phi  + {\cal C} \right)
\end{eqnarray}
 where we have introduced ${\bf \tilde \nabla }= \pmatrix{ \partial_{x^\prime}
\cr
 						\partial_{y^\prime} \cr
						 \partial_{z^\prime} }. $

It can be easily seen on the equations for $\rho $ and $h$ that the
change of variables
\be
  \tilde \rho( \bm{r^\prime} (t), t) = \rho( \bm{r^\prime}(t), t)\ \vert {\cal
Q}(t) \vert
\ee
and
\be
  \tilde h( \bm{r^\prime} (t), t) = h( \bm{r^\prime}(t), t)\ \vert {\cal Q}(t)
\vert^{\gamma-1}
\ee
will absorb the extra terms in the equations \ref{eqr} and \ref{eqh}.
The relevant variables we used are thus the components of the velocity
relative to the moving grid $\bf{ \tilde v}$, the scaled density $\tilde
\rho$ and the scaled enthalpy $ \tilde h.$
 The continuity and energy equations simplify further:
\be
  \partial_{\tau} \tilde \rho = - {\bf \tilde \nabla \cdot}
		\left( \tilde \rho {\bf \tilde v} \right) \label {eq. rhoc}
\ee
\be
\partial_{\tau}  \tilde h  =  -{\bf \tilde v \cdot \tilde \nabla} \tilde h -
			(\gamma -1) h  \left(
				{\bf \tilde \nabla \cdot  \tilde v} \right)
		\label {eq. hc}
\ee
and the Euler equation becomes:
\begin{eqnarray}
\partial_{\tau} {\bf \tilde v = - \tilde v \cdot \tilde \nabla \tilde v}
			&-& 2{\cal Q}^{-1} \dot{\cal Q} {\bf \tilde v}
			- {\cal Q}^{-1} \ddot{\cal Q} X^\prime \\ \nonumber
			&-& {\cal Q}^{-1}~^t{\cal Q}^{-1}
			{\bf \tilde\nabla }\left( {\tilde h  \over \vert{\cal Q}\vert^{\gamma-1}}
			+   \Phi  + {\cal C} \right) \ .
		\label {eq. vtilde}
\end{eqnarray}

We perform the calculations in a ``pseudo-spherical'' coordinate system
linked to  $X^\prime$ by the usual transformations
 		$ x^\prime = r^\prime \sin\theta^\prime \cos\phi^\prime $,
 		$ y^\prime = r^\prime \sin\theta^\prime \sin\phi^\prime $ and
 		$ z^\prime = r^\prime \cos\theta^\prime  $
because the relevant topology here is the one of a sphere. To be
complete, we need to add the Poisson equation, the equation of motion
of the star and the explicit form of the tidal potential. Finally, to
close the system of equations, we need to add a second order equation
of motion for the 9 coefficients of the matrix $\cal Q$. These last
equations, whose choice is a priori arbitrary, will be set up in such a
way that the grid motion is as close as possible to the mean matter
motion.

Many information can be drawn from the previous equations. The
transformations we made were inspired of course by the affine model.
Notice that if the motion of the fluid is exactly described by the
equations of motion of the affine star model and if we give to the
matrix $\cal Q$ the equation of motion of the deformation matrix $\cal
D$ (see section 2.1), the velocity of the matter with respect to an
inertial frame linked to the centre of mass of the star is exactly
given by
\be
 \bm{ v} = \dot{\cal Q} \hat{\bm{ r}} .
\ee
Hence, the vector $\bm{ r^\prime}$ is a constant and the relative speed
$\bf{ \tilde v}$ satisfies
\be
	\bf{ \tilde v  =  0}\ .
\ee
Moreover,  $\tilde \rho$ and $\tilde h$ are then constant and the scalar
fields $\rho$ and $h$ vary like powers of $\vert {\cal Q} \vert :$
\be
   \rho({ \bm r^\prime}(t), t) = \vert {\cal Q}(t) \vert^{-1}
  \rho( \bm{ r^\prime}(t=0), 0)
\ee
and
\be
   h( \bm{r^\prime}(t), t) = \vert {\cal Q}(t) \vert^{1 - \gamma}
  h( \bm{ r^\prime}(t=0), 0)\ .
\ee
One possible choice for the equation of motion of $\cal Q$ would be to
integrate simultaneoulsy the canonical affine equations for the
coefficients of the matrix $\cal D$ and the hydrodynamical equations.
However, that way, the relative velocity of the fluid with respect to
the moving grid at the surface of the star in the new frame would
increase as the affine model becomes less and less realistic. We would
not achieve a real improvement. We chose another way which has a number
of great numerical advantages: we minimize the modulus of the velocity
on the boundary of the star. The details of the  procedure is given in
the next Section. Doing so, we ensure that the star is approximately at
rest during the first two phases of its track around  the black hole,
where its deformation is roughly ellipsoidal, and that the surface of
the star is roughly given by $ \bm{ r^\prime = 1}$ in the moving frame.
We thus save computation time and have better precision.

\section {Numerical procedure}

Several tricks are used to reduce considerably the cost of the
calculations.

First, we take into account the symmetries of the problem. The
quadratic nature of all the potentials (tidal as well as gravitational)
makes it possible to reduce   the integration domain. Taking advantage
of the invariance with respect to the transformation:
\be \label{eqsym1}
(x,y) \rightarrow (-x,-y)
\ee
allows to make the calculation only on half a sphere. Moreover, this
problem is invariant   under reflection with respect to the orbital
plane
\be \label{eqsym2}
	z \rightarrow -z \ .
\ee
The required integration domain is therefore only a fourth of a sphere.

Second, the use of a well adapted moving grid has an important
influence on the numerics. If the motion of the grid is chosen to be as
close as possible to the mean motion of the fluid (which is always
possible in the case of tidal interactions), the computed components of
the velocity field almost vanish. As a consequence, the relative
velocity is very small, making the Courant condition (intrinsic to
every hydrodynamical problem)   very loose and hence allowing much
larger time steps than otherwise required (by several orders of
magnitude).

Let us mention that, because the transformation giving $X^\prime (t)$
from $X$ is linear, the pseudo-singularities due to the use of the
pseudo-spherical coordinate system $(r^\prime =0, \sin\theta^\prime
=0)$ are of the same kind than the usual one. Therefore they can be
handled like in Bonazzola and Marck (1990).

As mentionned above, we use the pseudo-spectral method to solve our set
of hydrodynamical equations. The method is a generalisation of the one
described with extensive details in Bonazzola and Marck (1990) which
takes into account the symmetries (eqs. \ref{eqsym1} and \ref{eqsym2}).  The
main advantage is that extremely high accuracy can be achieved without
using any kind of artificial viscosity. Our integration scheme is
second order and semi-implicit in time.

We used non dimensional variables and the non dimensioning procedure is
as follows.  Velocities are expressed in terms of the speed of light,
parameters of the orbits in terms of the Roche radius, positions of
fluid elements within the star in terms of the initial star radius.

Let us give now a sketch of the integration algorithm.
We first integrate explicitely all the terms of the right-hand side of
eq. \ref{eq. vtilde}, except those containing $\rm \ddot q _{ij}.$ Then we
calculate
 the  $\rm \ddot q _{ij}$ in order to minimize the quantity:

$$ \Sigma_{\rm boundary ~points} {\left(\rm {\bf v} -
dt {\cal Q}(t)^{-1}\ddot{\cal Q}(t){\bf  x}^\prime\right)}^2$$

We then get a linear system, the solution of which gives the relevant
 $\rm \ddot{q}_{ij}.$ They are
 integrated in time with a second order scheme to get
the $\rm \dot q_{ij}$  and the $\rm q_{ij}$ themselves.
  The initial conditions are provided by the canonical affine model.

Finally we perform the implicit phase of the integration.

This gives the following scheme:
$$\rm
{\bf v } \rightarrow {\bf v }^J + dt \left( c1~S_{\bf v}^J - c2 ~S_{\bf
v}^{J-1} \right)
$$
and
$$\rm
h \rightarrow h^J + dt \left( c1~S_{h}^J - c2 ~S_{h}^{J-1}\right)
$$
where
$$\rm
c1 = \left( 0.5 dt^J + dt^{J-1} \right)/dt^{J-1}
$$
$$\rm
c2 =  0.5 dt^J /dt^{J-1}
$$
and S stands for the source terms without the grid acceleration.

Then, computation of the $\ddot q_{ij}$ by the method described before.
Explicit integration of the remaining terms including $\ddot q_{ij}$ (only for
{\bf v}).
$$\rm
{\bf v } \rightarrow {\bf v } - dt^J {\cal Q}^{-1}\ddot{\cal Q} {\bf x}^\prime
$$
Finally, implicitation of the advection terms following the procedure:

$$
{{{\bf v }^{J+1} -{\bf v }}\over{dt^J}} = c r \partial_r {\bf v }^{J+1} - c r
\partial_r {\bf v }
$$
and the same for {\bf v}, where c is an adjustable parameter.

\bigskip
\section {A typical run}
\subsection{Run characteristics}

\begin{figure*}
\epsfysize=14.4cm
\epsfbox{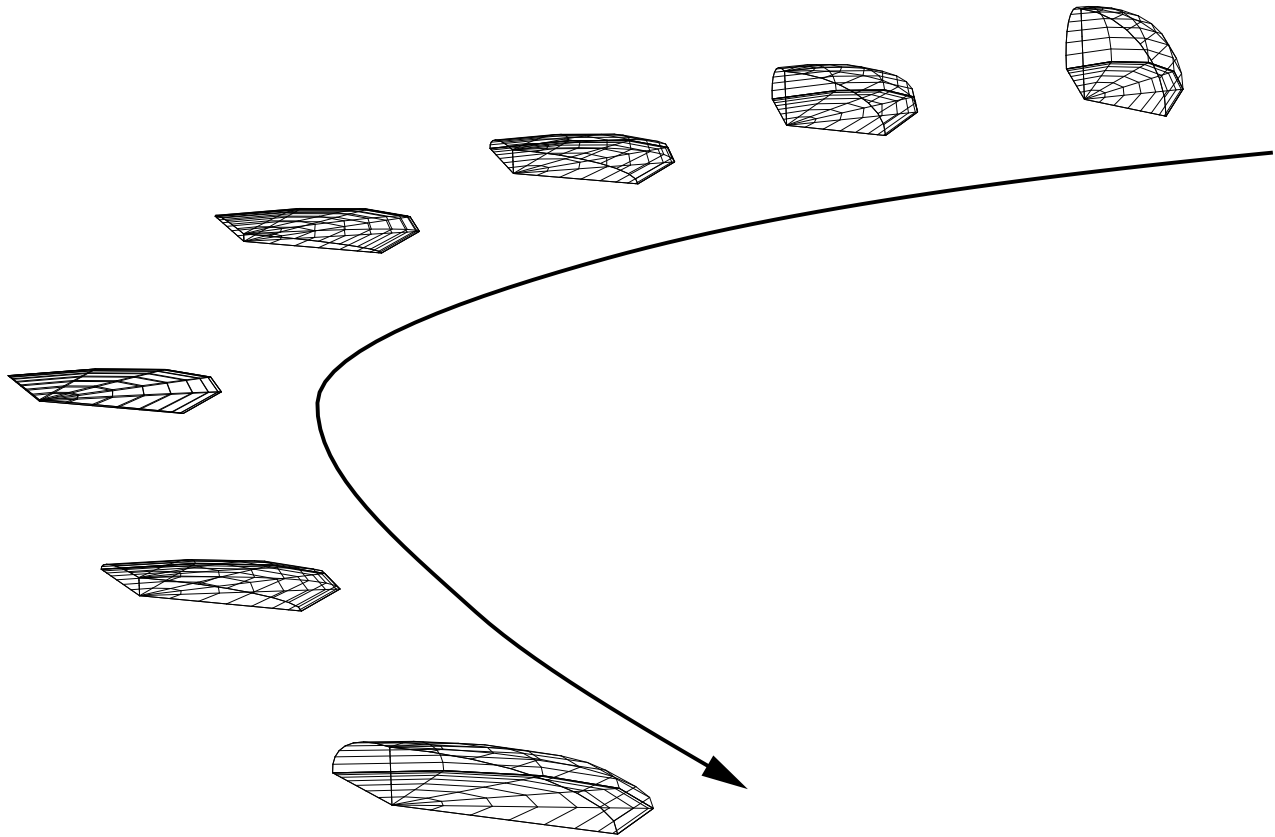}
\caption[ ]{Shape of the grid at different times. The grid is initially a
fourth of a sphere. Each individual plot is centered on the corresponding
location of the center of mass of the star on the track. Here $\beta=1.5$.
One clearly sees the flattening in the z direction and the rotation of the
principle axes in the orbital plane. }
\label{fig. Track}
\end{figure*}

We present here the result of one specific run. This is to be followed
in a next paper by an extensive study of the physics of the star core
during the encounter, with particular attention given to the possible
development of shocks and detonation of nuclear reactions. It is not
presently our aim to study the disruption and fate of the debris.

Since the curvature around a super-massive black hole is small, and
that the dimensions of a star are much smaller than the typical local
curvature radius, a newtonian treatment of the hydrodynamics is
justified.  However, a relativistic treatment of the orbit and the
tidal tensor is necessary for close encounters since they are known to
have cumulative effects along the track (Luminet and Marck 1985).

We take a polytropic star initially in spherical hydrostatic
equilibrium. We suppose that the orbit is parabolic and that the
deformation of the star is exactly described by the model of Carter and
Luminet down to the Roche radius. We thus integrate the canonical
affine model from r=5 (where tidal effects are negligible and the
spherical approximation fairly good) to r=1. When the star crosses the
Roche radius, we start the full hydrodynamical calculation, with the
initial conditions given by the affine model. The equation of state is
polytropic in this first step, with $\gamma = 5/3$.  The
gravitational potential is not treated in the exact way by solving at
each step the Poisson equation. We use instead a rough approximation
which consists to keep the potential constant in time, its value being
given by the initial model.  We expect this approximation to be
reasonable up to the point where autogravitation is definitely negligeable,
near the pericenter.

The star is characterized by the density contrast between the center
and the boundary, which we chose to be 10. The resolution of the
calculation is 17 modes in the radial direction, 9 modes in $\theta$
and 8 modes in $\phi$. For a specific encounter, we must specify a
penetration factor $\beta$.  The interesting range for $\beta$ is from
a few to 10 or more, since, for such values, extremely high
compressions are expected (Carter and Luminet 1983) and possible strong
electromagnetic effects may occur (Carter 1992).

The main difficulty in achieving numerical simulations of such
encounters, is that the density contrast between the equilibrium
configuration and the state of maximum compression may be as high as a
hundred or even more, according to the prediction of the affine model:
\be \label {eqrhomax}
{{\rho_{max}}\over{\rho_0}}  \sim \beta^{{2}\over{\gamma -1}}
	= \beta^{3} ~{\rm if}~\gamma=5/3
\ee
 The code must resist this compression phase.

The preliminary run we present here has $\beta = 1.5,$ corresponding to
$\eta \sim 0.54$ in the definition of Khokhlov et al.. Taking such values
for $\beta$ makes the approximation of a parabolic newtonian orbit very safe.
But even this relatively moderate penetration factor leads to a strong
deformation of the grid, as illustrated on fig \ref {fig. Track}.

\subsection{Comparison with the affine model}
\begin{figure}
\epsfxsize=8.8cm
\epsfysize=6.2cm
\epsfbox{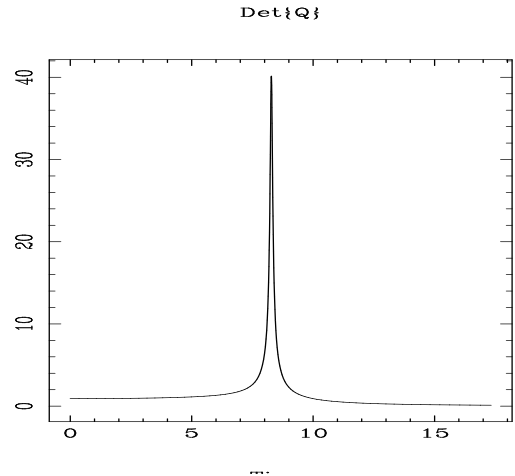}
 \caption[ ]{Plot of det {\cal Q} with respect to time. This picture sketches
the behaviour of the central density very well, due to eq \ref {eq. rhoc}. The
maximum compression rate occurs at the maximum of det {\cal Q}}
 \label{fig. DetQ}
\end{figure}

The important parameter in the affine model is the central density. More
precisely, the affine model makes a prediction on the maximum compression
rate, (see eq. \ref {eqrhomax}) which can be compared in this
hydrodynamical simulation with the ratio of the maximum to the initial
central density. This is important if one wants to speculate on the possible
enhancement of nuclear reactions. Previous numerical investigations of this
problem by Bicknell and Gingold (1983) found a milder dependance on $\beta$
than expected by the affine model, namely
${{\rho_{max}} / {\rho_0}}  \sim \beta^{1.5},$ even for moderate values
of $\beta.$ But at the same time, they find that a strong shock is formed
and reverses the collapse of the stellar material toward the  orbital plane.
Of course, such a complex behaviour is not contained in the affine model
and this might explain strong deviations for the value of the maximum
compression. On the other hand, one could also argue that their method
might not be reliable for very high compression and deformation, which can
also lead to significant  discrepancy. The numerical results by
Khokhlov et al. (1993a,b) deal only with low values of $\beta$
where essentially no compression occurs at all, except for one run,
making comparisons difficult.  Laguna et al. (1993), again using SPH,
end up roughly with the same $\beta$ dependance  as  Bicknell and
Gingold (1983). Here we do not give any dependance of the compression
rate with respect to the $\beta$ parameter since we have only one run.
A subsequent study including many runs and a careful study of the
dependance on $\beta$ of the results is still to come. However, this
first case seem to indicate a stronger compression than previously
found in numerical calculations (see Fig. \ref {fig. DetQ}).

 \begin{figure}
\epsfxsize=8.8cm
\epsfysize=6.2cm
\epsfbox{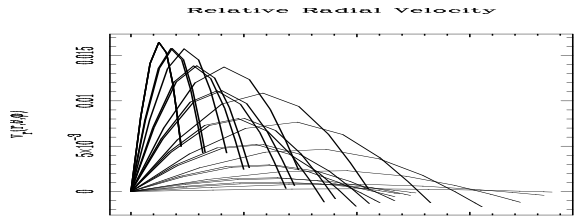}
 \caption[ ]{Plot of the radial relative velocity (actually computed by the
code)
as a function of $r$ for several values of the angles $\theta$ and $\phi$ when
the star reaches its periastron.}
 \label{fig. vr}
\end{figure}

\begin{figure}
\epsfxsize=8.8cm
\epsfysize=6.2cm
\epsfbox{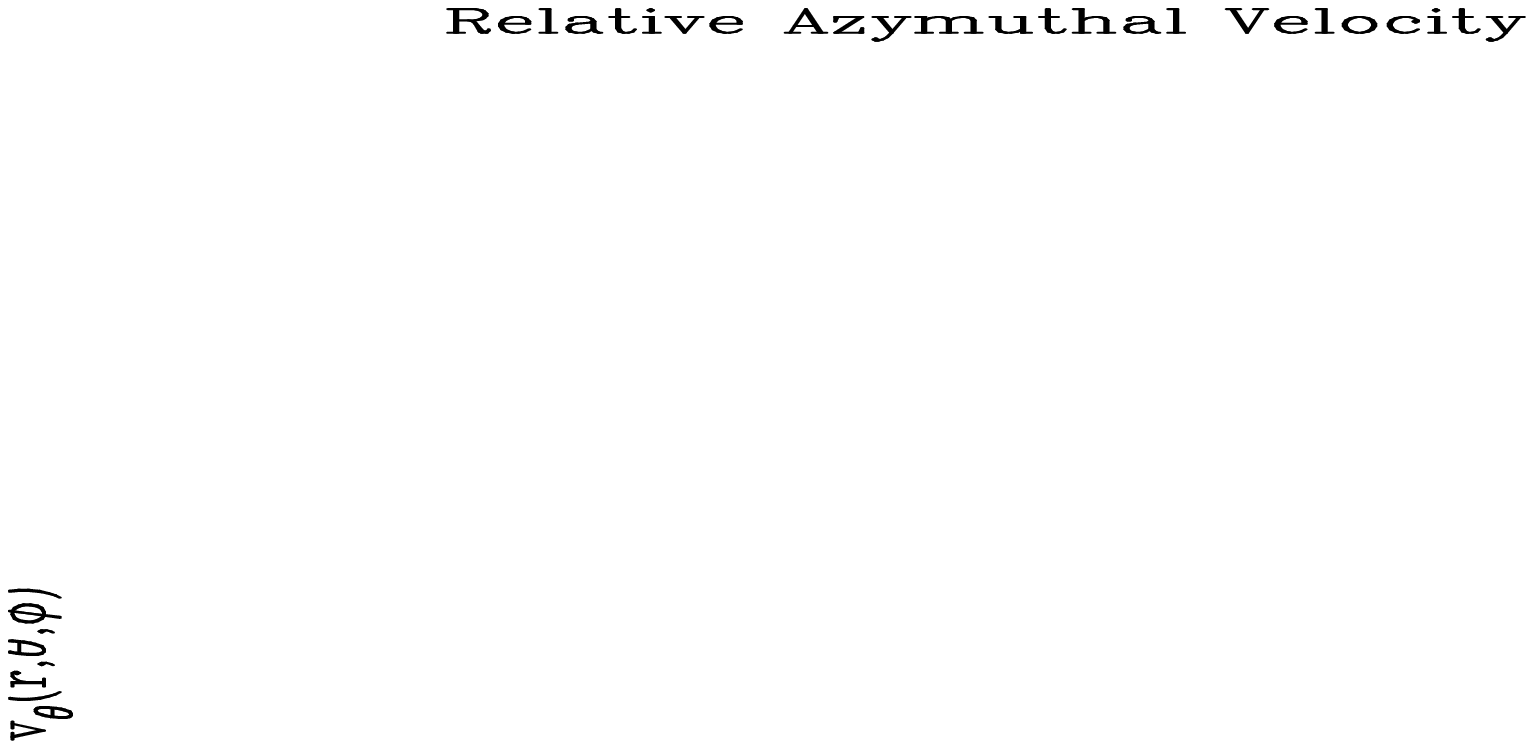}
 \caption[ ] {Plot of the relative velocity component $v_\theta$ (actually
computed by the code)
as a function of $r$ for several values of the angles $\theta$ and $\phi$ when
the star reaches its periastron. }
 \label{fig. vt}
\end{figure}

\begin{figure}
\epsfxsize=8.8cm
\epsfysize=6.2cm
\epsfbox{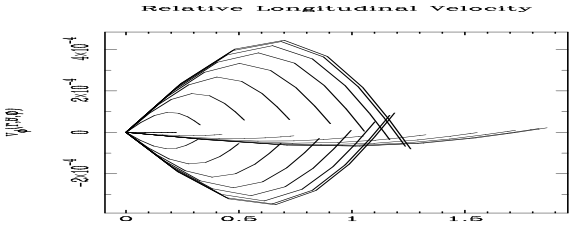}
 \caption[ ] {Plot of the relative velocity component $v_\phi$ (actually
computed by the code)
as a function of $r$ for several values of the angles $\theta$ and $\phi$ when
the star reaches its periastron.}
 \label{fig. vp}
\end{figure}

As expected, the use of a moving grid leads to much lower velocities
than with a static grid, at least in the first stages of the encounter,
as we could check by running the same calculation twice, once with a
moving grid according to section 3 and once with a static grid. As the
star approaches the periastron, it is very distorted and its internal
motions do not coincide any more with what one can expect from an
elliptical model. We present on figs \ref {fig.  vr}, \ref {fig. vt}
and \ref {fig. vp} the velocities relative to the moving grid, actually
calculated by the code, at the periastron.  Those velocities are the
deviations to the pure affine behaviour. This presentation, although
unusual, has the advantage of giving a feeling of the accuracy of the
affine model. A final point we would like to make is that those
relative velocities are plotted against the {\it real} radius, which
gives the information about the geometry of star. The reader should
bear in mind that the actual numerical calculation is made with an
elliptical grid, so that in every direction the radius ranges from 0 to
1.

\begin{figure}
\epsfxsize=8.8cm
\epsfysize=6.2cm
\epsfbox{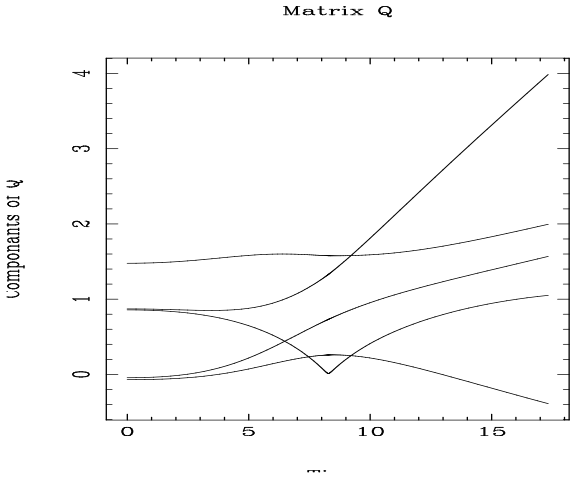}
 \caption[ ]{Plot of the five components of the matrix $\cal Q$ with time.
The overall behaviour is qualitatively identical to what one expects from the
affine model.  Time is given in non-dimensional units, scaled by $c/R_R$}
 \label{fig. Q}
\end{figure}

\begin{figure}
\epsfxsize=8.8cm
\epsfysize=6.2cm
\epsfbox{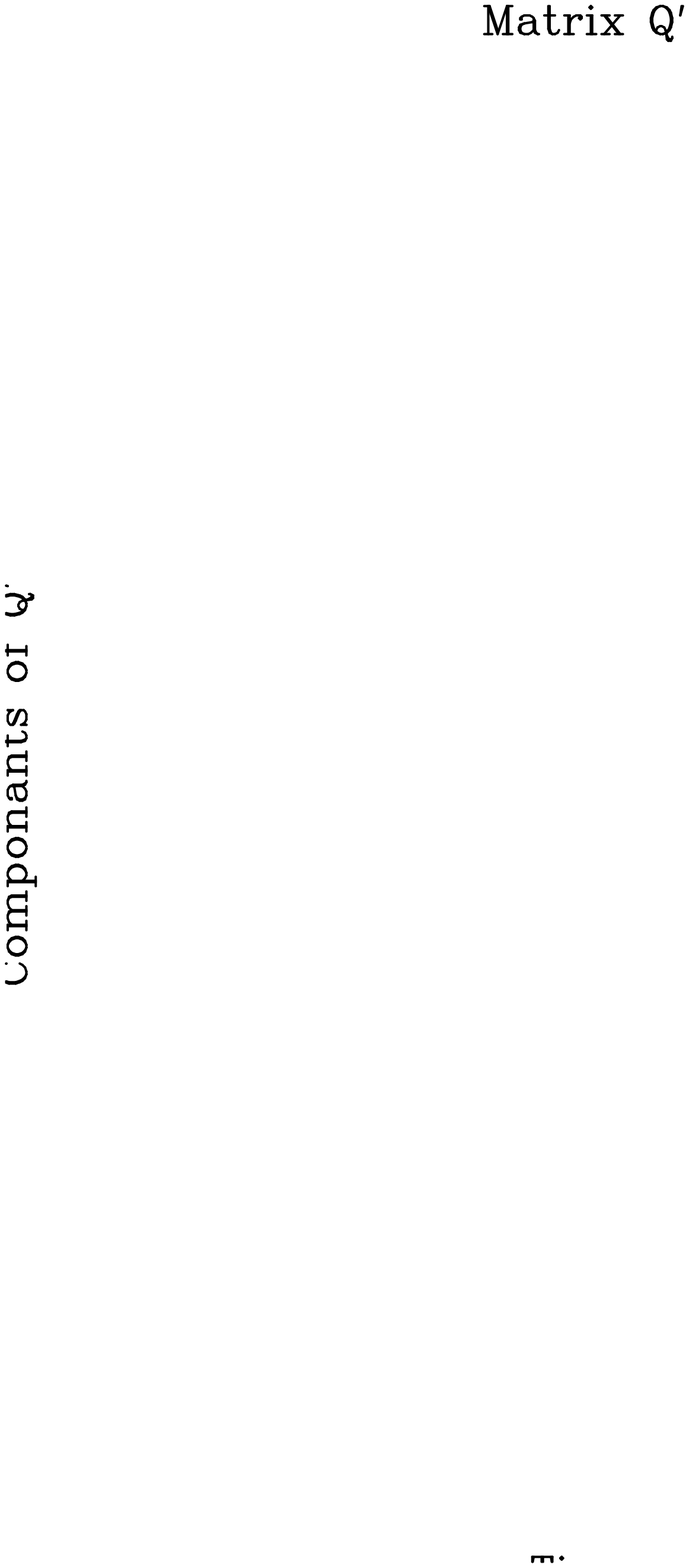}
 \caption[ ] {Same as fig \ref {fig. Q}, but for the matrix $\dot{\cal Q}$.  }
 \label{fig. Qp}
\end{figure}

\begin{figure}
\epsfxsize=8.8cm
\epsfysize=6.2cm
\epsfbox{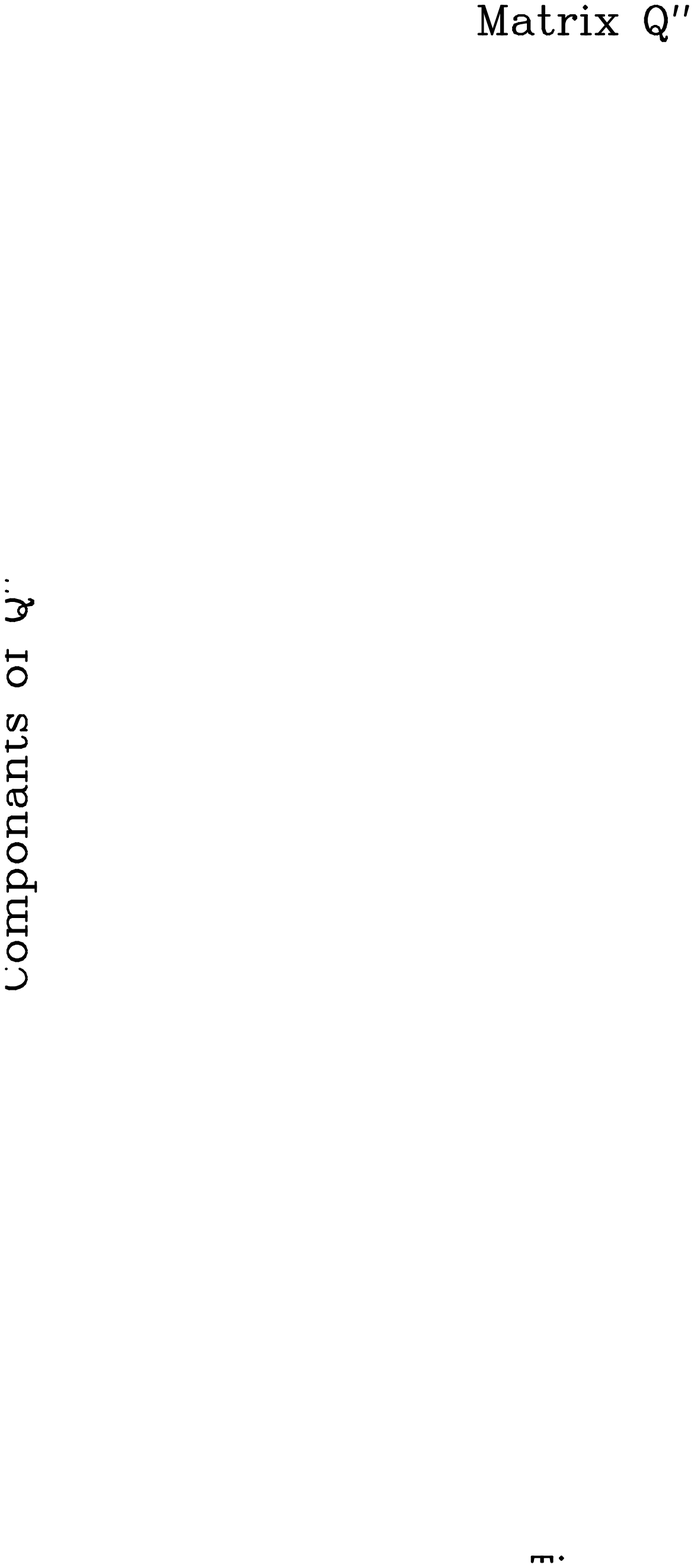}
 \caption[ ] {Same as fig \ref {fig. Q}, but for the matrix $\ddot{\cal Q}$.  }
 \label{fig. Qpp}
\end{figure}

The time evolution of the matrix elements $q_{ij}$ and their first and
second derivatives as well as $\rm det {\cal Q}$ are displayed on figs
\ref {fig. Q}, \ref {fig. Qp}, \ref {fig. Qpp} and \ref {fig. DetQ}.

\subsection{Comparison with other methods}

We already extensively discussed recent numerical results on the
same problem. We would like to give in this section a few more
technical elements, leading to quantitative comparison.

The advantage of the SPH method is its versatility. Once the code is
written, it can handle fairly easily almost any penetration factor. It
seems also to be quick enough so that the number of particles is not a
serious limitation. It seems possible within reasonable computing times
to triple the number of particles. At least, the results of Laguna et
al. do not seem to depend very much on the number of particles. Thus,
one can have with this method a very good overview of the problem
quickly. Moreover, for this particular problem, the study of the debris
is well treated by SPH, even if qualitative results can be already
given by a simple analysis of the geodesics. However, although it can
in principle handle shocks as well, the precision is probably
questionable, even if test cases are reproduced quite well and one has
to be cautious with the quantitative results, especially in the high
compression phase.

The eulerian method used by Khokhlov et al. is more suited for distant
encounters. It has the advantage of allowing a simple handling of
boundary conditions and is in general less delicate than the spectral
methods. But it is much more time consuming
and fails to treat high-$\beta$ encouters accurately because
the strong compression requires a prohibitive resolution.

Our approach is much more accurate than any of the others. Boundary
conditions and numerical implementation in general are relatively
complex but are now well mastered. We are able, with a moderate
resolution, to compute a quite close encouter. Thus, the computing time
is much less than for eulerian methods, owing also to the proper choice
of the grid and frame we made. The time required for the calculation
presented is typically 2.5 hours on a Silicon Graphics Indigo
workstation. Moreover, numerical difficulties grow very slowly with
increasing $\beta$, contrary to others.

\section{Discussion and conclusion}

We have presented a new numerical approach to the tidal interaction of
a star and a massive black hole, solving the hydrodynamical equtions in
a moving grid with spectral methods, which proves very powerful and
reliable.  Our moving grid allows us to maintain a constant resolution
within the star, whatever the deformation may be. By taking advantage
of most of the known analytical results to reduce the computation time
and complexity, we achieve fully three dimensional calculations within
a reasonable computing time, lower than the eulerian methods.
 Moreover, this method has no artificial viscosity and thus virtually
 no numerical dissipation, contrary to all other known methods (Laguna
et al. 1993, Khokhlov et al. 1993a,b).  We ran several
calculations with moderate value of the penetration factor, but still
greater than 1.

We expect that the forthcoming extensive study is likely to give
accurate additional results on the evolution of a star in a strong
tidal field. In particular, the detailed balance of energy transfer
could be examined as well as the oscillations subsequent to maximum
compression in the case of low $\beta$ encounters. For closer
encounters, the
 possible occurrence of shocks and the dependance of the maximum
density with $\beta$ will be checked.

However, the disruption of the star and the fate of the debris are
unlikely to be accurately treated by this code. SPH for example seems
to be much more adapted, but after all, the crescent shape of the
debris found by Laguna et al. seems to be quite easy to predict by a
simple geodesic calculation.

Finally, this method might be applicable to a number of other physical
applications including the oscillations of rotating stars.
\bigskip

\acknowledgements{We would like to thank Brandon Carter and Jean-Pierre Lasota
for stimulating our
interest on this subject and for fruitful discussions in the course of this
work. }
\onecolumn
\def \cp {\cos\phi}
\def \ct {\cos\theta}
\def \sp {\sin\phi}
\def \st {\sin\theta}
\def \cpd {\cos^2\phi}
\def \spd {\sin^2\phi}
\def \ctd {\cos^2\theta}
\def \std {\sin^2\theta}
\def \cpsp {\cos\phi \sin\phi}
\def \ctst {\cos\theta \sin\theta}
\def \quu {q_{11}}
\def \qud {q_{12}}
\def \qdu {q_{21}}
\def \qdd {q_{22}}
\def \qtt {q_{33}}
\def \quup {\dot q_{11}}
\def \qudp {\dot q_{12}}
\def \qdup {\dot q_{21}}
\def \qddp {\dot q_{22}}
\def \qttp {\dot q_{33}}
\def \quupp {\ddot q_{11}}
\def \qudpp {\ddot q_{12}}
\def \qdupp {\ddot q_{21}}
\def \qddpp {\ddot q_{22}}
\def \qttpp {\ddot q_{33}}
\def \detq {\vert {\cal Q}\vert}

\appendix
\section{Appendix}

In this appendix, we give the explicit form of the equations derived in Section
2 in pseudo-spherical coordinates. Recall that the continuity equation is
useless in the case of a polytropic star. We do not write here the energy
equation because it has the usual form. We drop the $\tilde {\ }$ and primes in
this appendix to make things simpler to read. The symbol $\Psi$ stands for all
the potentials in the Euler equation, that is, with the notations of Section 2:
\be
	 \Psi = {\tilde h \over \detq^{\gamma-1}} + \Phi + {\cal C}
\ee
where  $\Phi$ itself stands for the gravitational potential and $\cal C$ for
the
 tidal potential. This latter potential has the usual expression in terms of
the tidal
tensor ${\cal C}_{ij},$ in the newtonian approximation:

\be
	 {\cal C} = {\cal C}_{ij} { x_i x_j}
\ee
and
\be
{\cal C}_{ij} = { {G M_{h}}\over{r^5}} \left(  3 x_i x_j - r^2 \delta_{ij}
 \right).
\ee

The following equations are basically the components on a pseudo-spherical
frame of equation \ref {eq. vtilde} in the text. We applied the coordinate
transformation
followed by a projection onto the right vector basis.

\begin{eqnarray}
  \partial_{\tau } v_r = &-& v_r \partial_r v_r - (v_{\theta}/r)
	 \partial_{\theta}v_r
	 - v_{\phi}/(r \st) \partial_{\phi} v_r + (v_{\theta}^2 + v_{\phi}^2)/r
\nonumber \\
 	&+& 2  v_r\left(\std A - \ctd \qttp /\qtt \right)
		+ 2 \ctst v_{\theta} \left(A + \qttp / \qtt  \right)
	 + 2 \st\ v_{\phi} B	\nonumber \\
	 &+& \left( \std E -\ctd/\qtt^2 \right) \partial_r \Psi
		 +\ctst \left( 1 /\qtt^2  + E \right)\partial_{\theta}\Psi
		 + \st F \partial_{\phi} \Psi
		+ \left( \std H - \ctd \qttpp /\qtt \right) r
\end{eqnarray}

\begin{eqnarray}
 \partial_{\tau } v_{\theta } = &-& v_r \partial_r v_{\theta}
	- (v_{\theta }/r) \partial_{\theta}v_{\theta} - v_{\phi}/(r \st)
\partial_{\phi}
	v_{\theta} -  v_{\phi}^2 \cot\theta/r - (v_r v_{\theta})/r \nonumber \\
  &+& 2 \ctst v_r  \left( \qttp / \qtt - A \right)
	+ 2 v_{\theta}\left( \ctd A  - \std \qttp / \qtt \right) + 2 \ct v_{\phi} B
\nonumber \\
 &+& \ctst \left( 1/\qtt^2 + E \right) \partial_r \Psi
 +      \left( E \ctd  - \std /\qtt^2 \right) \partial_{\theta} \Psi
 + \ct F \partial_{\phi} \Psi
+ \ctst r  \left( \qttpp /\qtt + H \right)
\end{eqnarray}
\begin{eqnarray}
 \partial_{\tau } v_{\phi } = &-& v_r \partial_r v_{\phi} -
	(v_{\theta }/r )\partial_{\theta}v_{\phi}
	- v_{\phi}/(r \st) \partial_{\phi} v_{\phi}
	- (v_r v_{\phi})/r - v_{\phi}v_{\theta } \cot\theta/r	\nonumber \\
& +&  2 C \left( \st v_r + \ct v_{\theta} \right) + 2 v_{\phi} D \nonumber 	\\
   &+& F \left( \st \partial_r \Psi + \ct \partial_{\theta} \Psi \right)
	- G  \partial_{\phi} \Psi + \st I r
\end{eqnarray}
where we adopted the following shorthand notations:
\be
A =
  \left(\cpsp(\quup\qdu -\quu\qdup -\qudp\qdd +\qud\qddp )
	+\cpd(\qud\qdup-\quup\qdd )+\spd(\qudp\qdu-\quu\qddp)\right) /\delta
\ee
\be
B =
  \left(\cpsp(\quup\qdd-\quu\qddp+\qudp\qdu-\qud\qdup)
  +\cpd(\qud\qddp-\qudp\qdd) + \spd (\quu\qdup-\quup\qdu)\right) /\delta
\ee
\be
C =
  \left(\cpsp (\quup\qdd -\quu\qddp+\qudp\qdu-\qud\qdup )
	+\cpd(\quup\qdu-\quu\qdup )+\spd(\qudp\qdd-\qud\qddp)\right) /\delta
\ee
\be
D =
  \left(\cpsp(\quu\qdup-\quup\qdu-\qud\qddp+\qudp\qdd)
  +\cpd(\qudp\qdu-\quu\qddp) + \spd (\qud\qdup-\quup\qdd)\right) /\delta
\ee
\be
E=
 \left(2\cpsp(\quu\qud+\qdu\qdd)-\cpd(\qud^2+\qdd^2) -
\spd (\quu^2 + \qdu^2) \right) /\delta^2
\ee
\be
F =
  \left(\cpsp(\qud^2 -\quu^2 - \qdu^2 + \qdd^2) +
(\cpd-\spd)(\quu\qud+\qdu\qdd) \right) /\delta^2
\ee
\be
G  =
  \left(2\cpsp(\quu\qud +\qdu\qdd) +
\cpd(\quu^2 +\qdu^2) + \spd(\qud^2+\qdd^2)\right)  /\delta^2
\ee
\be
H =
 \left(\cpsp(\quupp\qdu-\quu\qdupp-\qudpp\qdd+\qud\qddpp)
+\cpd(\qud\qdupp-\quupp\qdd )+\spd(\qudpp\qdu-\quu\qddpp)\right) /\delta
\ee
\be
I =
  \left(\cpsp(\qudpp\qdu-\qud\qdupp+\quupp\qdd-\quu\qddpp)
	+\cpd(\quupp\qdu-\quu\qdupp )+\spd(\qudpp\qdd-\qud\qddpp)\right) /\delta
\ee
and
\be
\delta = \quu\qdd - \qud\qdu \ .
\ee

\twocolumn

\end{document}